# Evidence for the Collective Nature of the Glass Transition

K. Moch, P. Münzner, R. Böhmer, C. Gainaru

*Fakultät Physik, Technische Universität Dortmund, 44221 Dortmund, Germany*

Self- and cross-correlation dynamics of deeply supercooled liquids were recently identified using photon correlation spectroscopy and spin relaxometry on the one hand, and using dielectric investigations on the other. These results fueled a controversial discussion whether the "generic" response identified by the former techniques or rather the non-universal dielectric response reflect the liquid's structural relaxation. The present study employs physical aging and oscillatory shear rheology to directly access the structural relaxation of a non-associating glass forming liquid and reveals that collective equilibrium fluctuations of simple liquids and not single-particle dynamics govern their structural relaxation.

A popular notion regarding glasses holds that they are liquids fallen out of time, i.e., structurally disordered solids with thermally arrested dynamics. The most salient features of the structural dynamics of glass-forming liquids can be summarized with reference to their non-exponential, non-linear, and non-Arrhenius nature – succinctly coined as the three "nons" [1]. Yet, the microscopic origin of these well-established, practically universally observed characteristics of supercooled liquids remains a source of fruitful controversy. The most important debate centers on the nature of the structural relaxation itself and explores whether or not vitrification can be regarded as a collective phenomenon. In viscous liquids, the coupling among structural subunits was considered early on [2] and ever since remained an inspiring idea in the search for evidence unraveling the collective nature of viscous liquid dynamics [3]. In this vein, indications for molecular cooperativity were reported based on nuclear magnetic resonance (NMR) investigations [4], confinement studies [5], and atomic force measurements of dielectric noise [6]. More recently, theoretical work [7] inspired nonlinear dielectric studies [8,9] that relate the emergence of "humps" in higher-order susceptibilities to a collective, multi-particle behavior, an assignment that is, however, not uncontested [10,11].

Several classes of glass-forming liquids exhibit well recognized collective relaxation modes already in their linear responses. Here, polymer melts come to mind, in which concerted motions within the macromolecular backbone, the well-known normal modes, relax slower than the structural, segmental modes [12]. A wealth of nonpolymeric glass formers, many of them featuring hydrogen bonds, also display clear spectral signatures of supramolecular relaxations. Prominent examples include the monohydroxy alcohols, for which the time scale of their so-called Debye-like process is sometimes more than 1000 times slower than their structural relaxation [13]. The collective nature of the main dielectric response of these materials has been rationalized with reference to the large contrast between the dielectric and NMR relaxometry time scales reflecting their cross-correlation (collective) and self-correlation (single-particle) molecular motions, respectively [14].

Even for much smaller spectral separations, and aided by photon correlation spectroscopy (PCS), a decomposition of the total response in terms of single- and multi-particle dynamics was demonstrated not only for the dielectric spectra of monohydroxy alcohols [15], but also for the prototypical H-bonded glass former glycerol [16]. The success of this decomposition rests on the observation that the main relaxation process probed by PCS (reflecting single-particle dynamics) is significantly faster than the dielectric relaxation and that the PCS spectra display a "generic shape" [17], which then necessarily is temperature independent. Avoiding the complexities inherent in the physics of hydrogen bonded liquids, this decomposition was recently demonstrated also for tributyl phosphate (TBP) [18], a moderately polar, non-associating liquid [19] featuring a calorimetric glass transition temperature $T_g$ near 141 K [20].

The finding of a "generic shape" for the main relaxation detected in PCS is corroborated by recent results from NMR relaxometry [21]. The quasi-universal spectral shape of the reorientational susceptibility thus identified for a large number of glass formers contrasts with the major variability of *dielectrically* detected spectral shapes or widths. These in turn were shown to correlate well with the dielectric relaxation strength $\Delta\varepsilon$ [22] and were rationalized in terms of changes in the intermolecular potential [23]. In the limit of sufficiently small $\Delta\varepsilon$, and clearly shown for only a handful of low-polarity non-associating liquids, the "generic shape" detected by NMR relaxometry or PCS was suggested to be approached by the dielectric spectra as well [18,21].

Thus, currently one faces the situation that in some simple (nonassociating and nonpolymeric) liquids, NMR relaxometry and PCS probe a "generic" dynamics which can be significantly different from that revealed by dielectric spectroscopy. The central question arises which of these methods, that probe different equilibrium fluctuations, most closely reflects the structural rearrangements within these materials. (i) Should, following recent suggestions [18], the structural relaxation be associated with the "generic" single-particle dynamics? Are the dielectrically detected molecular cross-correlations, hence, useful only to estimate the magnitude of the electric polarization in different materials? (ii) Or do the collective dielectric modes govern the time scale of the structural relaxation in polar liquids?

To address these questions, we study the nonassociating liquid TBP for which recent PCS and dielectric investigations, even close to $T_g$, revealed significantly different dynamics. Performing shear rheology and physical aging for TBP, we demonstrate that near $T_g$ its microscopic flow and its structural recovery persist on time scales significantly longer than those identified by PCS. Moreover, based on its equilibrium *dielectric* behavior, we successfully estimate the time scale of its nonlinear aging response which is highly relevant for the glassy behavior of this material.

Dielectric and PCS spectra of TBP, for $T = 140$ and 147 K taken from Ref. 18, are shown in Fig. 1. The shapes of the PCS spectra adapt themselves perfectly to the high-frequency parts of the dielectric losses. The dielectric spectrum at $T = 137$ K is from the present work; experimental details are given as supplementary information (SI). The inset of Fig. 1 illustrates that a Debye-type low-frequency contribution remains after subtracting the suitably normalized PCS susceptibility from the dielectric loss $\varepsilon''$, with the colored areas highlighting how the various contributions can give rise to the overall dielectric spectrum.

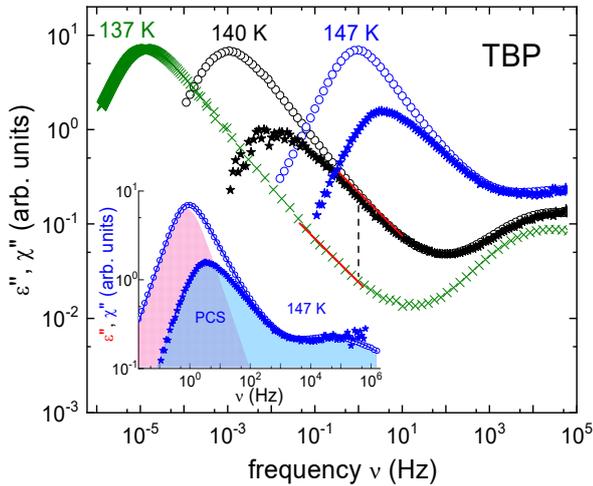

Fig. 1. The dielectric and PCS susceptibility spectra probed for TBP at 140 and 147 K are taken from ref. 18. The dielectric spectrum, recorded at 137 K after the sample was equilibrated at this temperature for about $3\times10^5$ s, is from the current work. The solid red lines indicate a power law, $\varepsilon'' \propto v^{-0.48}$ of the dielectric susceptibility near the aging detection frequency (1 Hz, as indicated by the vertical line). Note that at 140 K the PCS and dielectric susceptibility peaks are separated by a factor of ten. The inset, adapted from ref. 18, was used to suggest that the dielectric response of TBP exhibits an additional slow Debye process (highlighted in red) with respect the PCS susceptibility (highlighted in blue).

To decide among the two above scenarios, it is mandatory to properly identify the structural relaxation by experimental means that provide direct access to it. Calorimetry and viscometry are recognized as suitable probes for TBP [20,24]. In a first step, we measured its complex shear mechanical compliance, $J^* = J' - iJ''$, which represents a susceptibility function, similar to the dielectric and PCS functions considered in Fig. 1. Since the loss part of the compliance $J''(v)$ is severely superimposed by contributions from molecular flow, thus hampering a direct comparison with the dielectric and PCS spectra, in Fig. 2 we focus on the storage part of the compliance, $J'(v)$. For details see the SI where we also show that our viscoelastic results nicely agree with expectations based on high-$T$ viscosity and low-$T$ calorimetry investigations of TBP [24]. While in its fluid and moderately supercooled regime the dielectric and PCS time scales differ considerably from those of the enthalpic and rheological responses [20], close to $T_g$ the temperature dependence of all relaxation times is highly similar (see the SI) and, except for PCS, also their absolute values. We interpret this convergence of all (except the PCS) time scales to indicate that when $T_g$ is closely approached, the coupling among the various degrees of freedom generates a common cooperative behavior.

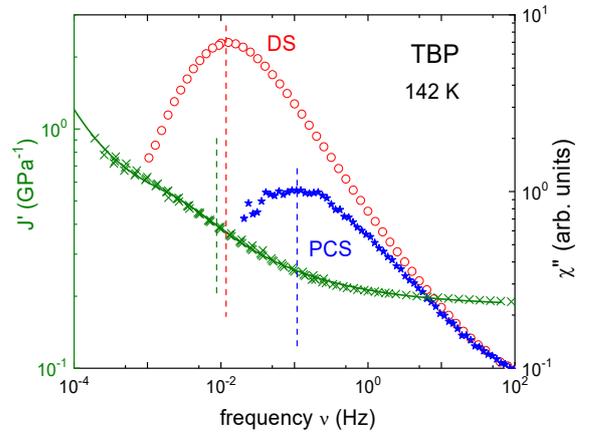

Fig. 2. Master plot of the shear susceptibility spectra of TBP. The reference temperature was chosen as 142 K in order to facilitate comparison with dielectric spectroscopy (DS) and PCS experiments carried out at the same temperature. The vertical lines indicate the characteristic frequencies of the different isothermally probed susceptibilities. Note that the peak separation between the PCS and dielectric susceptibility is about a factor of 10. The absence of a low-frequency plateau in $J'(v)$, corresponding to the recoverable compliance, is due to low-torque artifacts. The solid line reflects a fit with a Cole-Davidson function [25] augmented by a power law accounting for the latter.

From Fig. 2 it is obvious that the frequency position of the clearly resolved inflection point in $J'(v)$ is close to the characteristic one of the dielectric loss peak, but differs significantly from that of the PCS susceptibility, which at 142 K indicates an about ten times faster relaxation. Although, one may argue that comparisons of results from different techniques are prone to suffering from differences in the experimental conditions, temperature readings, and other circumstances, this observation nevertheless confirms that the dielectric, and not the light scattering experiment, provides direct access to the viscoelastically detected structural relaxation.

Then, how can the structural relaxation be monitored in a way that is independent of the specific probe? A suitable means is to exploit physical aging experiments. These are based on inducing a change of the liquid's structural state, e.g., by a temperature step, and to detect its subsequent evolution by means of techniques as diverse as neutron detected vibrational spectroscopy [26], dilatometry [27,28], shear-mode [29], and not the least dielectric spectroscopy [30,31]. Common to all these techniques is that the largely varying dynamics to which these methods are primarily sensitive are all driven by or intimately coupled to the same structural state within the liquid and its temporal evolution. A drawback of the aging diagnostics, by whatever technique, is that it is necessarily nonlinear in the sense that its outcome depends on direction and magnitude of the externally applied temperature perturbation. The consequences of this circumstance were recognized early on [27] and can adequately be captured within the well-known TMNH [32] or KAHR models [33].

To tackle the present issue, we thus aim at checking whether or not the dielectrically detected and rheologically confirmed time scales truly reflect that of the structural relaxation. Therefore, we performed equilibrium electric-field-step as well as a temperature-jump experiments within the *same* dielectric cell and setup, so as to minimize potential differences in experimental conditions. Fig. 3 presents the temporal evolution of the electrical polarization that was recorded after switching the external field isothermally at 137 K. These data were Fourier transformed, combined with those probed in the frequency domain at the same temperature, and included in Fig. 1. In Fig. 3 the *time* dependent polarization decay is shown in normalized form in order to facilitate comparison with the results from the aging experiment. Here, the temperature was stepped (more precisely rapidly ramped) from 140 K down to a base temperature of 137 K, see the temperature protocol depicted in the inset of Fig. 3. In order to track the subsequent structural changes, we recorded the dielectric loss at a frequency of 1 Hz. This detection frequency was chosen because, as one recognizes from Fig. 1, it lies within a moderately broad spectral range in which the dielectric loss follows an apparent power law. Under these provisos, this high-frequency monitoring is largely equivalent to probing the time evolution of the dielectric loss peak frequency [31].

The time dependent dielectric loss, $\varepsilon''(1\,\mathrm{Hz})$, of TBP measured while the sample ages at its base temperature, is shown in Fig. 3. One recognizes that this function decays faster than the dielectric polarization, thus at first glance supporting the implication from Ref. 18 that is illustrated in the inset of Fig. 1, which states that structural relaxation proceeds *faster* than inferred from the dielectric loss peak. However, based on the data in Fig. 3, such a conclusion must be considered premature because, after a thermal down-step, the sample initially relaxes with the rate corresponding to the starting temperature. Then, although self-retardation effects evolve in the course of physical aging, the resulting aging curve necessarily appears to be faster, typically by a factor of 2…3, as compared to the response measured in thermal equilibrium [34,35]. As the red line in Fig. 3 shows, the experimentally measured aging curve is compatible with calculations (for details see the SI) performed on the basis of the equilibrium dielectric response within the TMNH model.

PCS data are not available for $T$ = 137 K. By assuming the factor-of-ten separation valid at 140 K (Fig. 1) and 142 K (Fig. 2), from Fig. 3 we see that the PCS response thus expected at 137 K is considerably faster than the measured aging curve. Thus, at least for a liquid beyond the low-polarity limit, the aging experiment indeed provides unequivocal evidence that it is the collective *dielectric* response which probes the structural relaxation.

While this finding contrasts with recent claims [18], it is in harmony with the relevance of a coupling among the structural subunits in glass forming liquids [3] and may be linked with cross-correlation effects that prominently show up in dielectric spectroscopy [36,37].

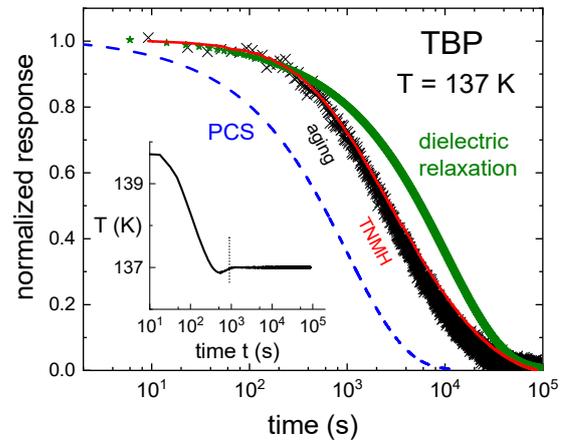

Fig. 3. The normalized equilibrium dielectric response of TBP probed at 137 K (greed solid stars) is compared with the normalized 1 Hz dielectric aging data (black crosses) probed at the same base temperature and the expectation of the equilibrium PCS response at 137 K (blue dashed line). The latter has been estimated assuming the same one decade for the time scale difference between the PCS and dielectric responses, obvious from the results presented in Fig. 1 (at 140 K) and Fig. 2 (at 142 K). The solid red line is calculated using the TMNH formalism, see the SI. The inset depicts the temperature variation prior and during the acquisition of the dielectric aging results. The vertical line indicates a possible choice for the starting condition in this experiment, another one is discussed in the SI.

To summarize, we performed oscillatory shear investigations and physical aging experiments on a simple glass forming liquid for which, on the basis of PCS and dielectric responses, single- and multi-particle dynamics, respectively, were previously distinguished. Our results demonstrate that close to $T_g$ the time scales of the molecular flow and of the structural recovery are governed by the same collective dynamics which controls the *dielectric* response of TBP. Stipulating that "a theory of the physics of glass formation should aim to explain in a unified manner" [38], prior to the present study the choice was either (i) to consider the non-universally shaped dielectric relaxation as the starting point of such theories or else (ii) to start from the single-particle relaxation identified, e.g., by PCS, with its universal shape that is in harmony with results from several other techniques. The present results favor choice (i), since they clearly demonstrate that based on cross-correlated equilibrium fluctuations, but not on self-correlations, one can predict the structural relaxation of simple liquids.

———————

[1] J. C. Dyre, Colloquium: The glass transition and elastic models of glass-forming liquids, Rev. Mod. Phys. **78**, 953 (2006).
[2] G. Adam and J. H. Gibbs, On the temperature dependence of cooperative relaxation properties in glass-forming liquids, J. Chem. Phys. **43**, 139 (1965).
[3] K. L. Ngai, *Relaxation and Diffusion in Complex Systems* (Springer, Berlin, 2011).
[4] U. Tracht, M. Wilhelm, A. Heuer, H. Feng, K. Schmidt-Rohr, and H. W. Spiess, Length scale of dynamic heterogeneity at the glass transition determined by multidimensional nuclear magnetic resonance. Phys. Rev. Lett. **81**, 2727 (1998); X. H. Qiu and M. D. Ediger, Length scale of dynamic heterogeneity in supercooled D-sorbitol: Comparison to model predictions. J. Phys. Chem. B **107**, 459 (2003).
[5] M. Arndt, R. Stannarius, H. Groothues, E. Hempel, and F. Kremer, Length Scale of Cooperativity in the Dynamic Glass Transition, Phys. Rev. Lett. **79**, 2077 (1997).
[6] E. Vidal Russell and N. E. Israeloff, Direct observation of molecular cooperativity near the glass transition, Nature **408**, 695 (2000).
[7] C. Dalle-Ferrier, C. Thibierge, C. Alba-Simionesco, L. Berthier, G. Biroli, J.-P. Bouchaud, F. Ladieu, D. L'Hôte, and G. Tarjus, Spatial correlations in the dynamics of glassforming liquids: Experimental determination of their temperature dependence, Phys. Rev. E **76**, 041510 (2007).
[8] S. Albert, T. Bauer, M. Michl, G. Biroli, J.-P. Bouchaud, A. Loidl, P. Lunkenheimer, R. Tourbot, C. Wiertel-Gasquet, and F. Ladieu, Science **352**, 1308 (2016); R. Casalini and C. M. Roland, Nonlinear dielectric spectroscopy of propylene carbonate derivatives, J. Chem. Phys. **148**, 134506 (2018).
[9] R. Richert (ed.), *Nonlinear Dielectric Spectroscopy*, (Springer, Cham, 2018).
[10] G. Diezemann, Higher-order correlation functions and nonlinear response functions in a Gaussian trap model, J. Chem. Phys. **138**, 12A505 (2013).
[11] P. Kim, A. R. Young-Gonzales, and R. Richert, Dynamics of glass-forming liquids. XX. Third harmonic experiments of non-linear dielectric effects versus a phenomenological model, J. Chem. Phys. **145**, 064510 (2016).
[12] G. Strobl, *The physics of polymers* (Springer, Berlin, 1997); K. Adachi and T. Kotaka, Dielectric normal-mode relaxation, Prog. Polym. Sci. **18**, 585 (1993).
[13] R. Böhmer, C. Gainaru, and R. Richert, Structure and dynamics of monohydroxy alcohols-milestones towards their microscopic understanding, 100 years after Debye, Phys. Rep. **545**, 125 (2014).
[14] A. Volmari and H. Weingärtner, Cross terms and Kirkwood factors in dielectric relaxation of pure liquids, J. Mol. Liq. **98-99**, 293 (2002); U. Kaatze, Dielectric and structural relaxation in water and some monohydric alcohols, J. Chem. Phys. **147**, 024502 (2017).
[15] J. Gabriel, F. Pabst, A. Helbling, T. Böhmer, and T. Blochowicz, Nature of the Debye-Process in Monohydroxy Alcohols: 5-Methyl-2-Hexanol Investigated by Depolarized Light Scattering and Dielectric Spectroscopy, Phys. Rev. Lett. **121**, 035501 (2018).
[16] J. P. Gabriel, P. Zourchang, F. Pabst, A. Helbling, P. Weigl, T. Böhmer, and T. Blochowicz, Intermolecular cross-correlations in the dielectric response of glycerol, Phys. Chem. Chem. Phys. **22**, 11644 (2020).
[17] F. Pabst, J. Gabriel, T. Böhmer, P. Weigl, A. Helbling, T. Richter, P. Zourchang, T. Walther, and T. Blochowicz, Universal Structural Relaxation in Supercooled Liquids, arXiv:2008.01021
[18] F. Pabst, A. Helbling, J. Gabriel, P. Weigl, and T. Blochowicz, Dipole-dipole correlations and the Debye process in the dielectric response of nonassociating glass forming liquids, Phys. Rev. E **102**, 010606(R) (2020).
[19] Here, the term "non-associating" implies the absence of specific *chemical* interactions including covalent, ionic, and hydrogen bonding, metal-ligand interaction, or $\pi$-$\pi$ stacking.
[20] M. K. Saini, Y. Gu, T. Wu, K. L. Ngai, and L.-M. Wang, Deviations of dynamic parameters characterizing enthalpic and dielectric relaxations in glass forming alkyl phosphates, J. Chem. Phys. **149**, 204505 (2018).
[21] T. Körber, R. Stäglich, C. Gainaru, R. Böhmer, and E. A. Rössler, Systematic differences in the relaxation stretching of polar molecular liquids probed by dielectric vs magnetic resonance and photon correlation spectroscopy, J. Chem. Phys. **153**, 124510 (2020).
[22] M. Paluch, J. Knapik, Z. Wojnarowska, A. Grzybowski, and K. L. Ngai, Universal Behavior of Dielectric Responses of Glass Formers: Role of Dipole-Dipole Interactions, Phys. Rev. Lett. **116**, 025702 (2016).
[23] P. Bordat, F. Affouard, M. Descamps, and K. L. Ngai, Does the Interaction Potential Determine Both the Fragility of a Liquid and the Vibrational Properties of its Glassy State?, Phys. Rev. Lett. **93**, 105502 (2004).
[24] T. Wu, X. Jin, M. K. Saini, Y. D. Liu, K. L. Ngai, and L. Wang, Presence of global and local α-relaxations in an alkyl phosphate glass former, J. Chem. Phys. **147**, 134501 (2017).
[25] D. W. Davidson and R. H. Cole, Dielectric Relaxation in Glycerol, Propylene Glycol, and n-Propanol, J. Chem. Phys. **19**, 1484 (1951).
[26] E. Duval, L. Saviot, L. David, S. Etienne, and J. F. Jal, Effect of physical aging on the low-frequency vibrational density of states of a glassy polymer, Europhys. Lett. **63**, 778 (2003).
[27] R. O. Davies and G. O. Jones Thermodynamic and kinetic properties of glasses, Adv. Phys. **2**, 370 (1953).
[28] A. J. Kovacs, La contraction isotherme du volume des polymères amorphes, J. Polym. Sci. **30**, 131 (1958).
[29] N. B. Olsen, J. C. Dyre, and T. Christensen, Structural relaxation monitored by instantaneous shear modulus, Phys. Rev. Lett. **81**, 1031 (1998).
[30] see, e.g., E. Schlosser and A. Schönhals, Dielectric relaxation during physical ageing, Polymer **32**, 2135 (1991); D. Cangialosi, M. Wübbenhorst, J. Groenewold, E. Mendes, and S. J. Picken, Diffusion mechanism for physical aging of polycarbonate far below the glass transition temperature studied by means of dielectric spectroscopy, J. Non-Cryst. Solids **351**, 2605 (2005).
[31] T. Hecksher, N. B. Olsen, K. Niss, and J. C. Dyre, Physical aging of molecular glasses studied by a device allowing for rapid thermal equilibration, J. Chem. Phys. **133**, 174514 (2010).
[32] I. M. Hodge, Enthalpy relaxation and recovery in amorphous materials, J. Non-Cryst. Solids **169**, 211 (1994) and references cited therein.
[33] A. J. Kovacs, J. J. Aklonis, J. M. Hutchinson, and A. R. Ramos, Isobaric volume and enthalpy recovery of glasses. II. A transparent multiparameter theory, J. Polym. Sci.: Polym. Phys. Ed. **17**, 1097 (1979).
[34] P. Lunkenheimer, R. Wehn, U. Schneider, and A. Loidl, Glassy Aging Dynamics, Phys. Rev. Lett. **95**, 055702 (2005); R. Richert, P. Lunkenheimer, S. Kastner, and A. Loidl, J. Phys. Chem. B **117**, 12689 (2013).
[35] T. Hecksher, N. B. Olsen, and J. C. Dyre, Communication: Direct tests of single-parameter aging, J. Chem. Phys. **142**, 241103 (2015).
[36] G. Williams, Use of the dipole correlation function in dielectric relaxation, Chem. Rev. **72**, 55 (1972).
[37] F. Kremer and A. Schönhals (Eds.), *Broadband Dielectric Spectroscopy* (Springer, Berlin, 2003).
[38] K. Niss and T. Hecksher, Perspective: Searching for simplicity rather than universality in glass-forming liquids, J. Chem. Phys. **149**, 230901 (2018).